# Analysis of Interference in Wireless Networks

Zeeshan Haider, Muhammad Saleem and T. Jamal

luckier19@gmail.com

PIEAS University Islamabad

ABSTRACT

As wireless systems grow rapidly worldwide, one of the most important things, wireless systems designers and service providers' face, is interference. Interference decreases coverage, capacity [1], and limits the effectiveness of both new and existing systems. It is very difficult to avoid because wireless communications systems must exist together in extremely complex signal environments. These environments are consisting of multiple operating wireless networks [2]. At the same instant, new technologies and signal sources in Wireless Local Area Networks (WLANs) and digital video broadcasting are jeopardized to wireless communications service. This article provides a survey and analysis of interference in Wireless Network.

1. INTRODUCTION

There are different types of interferences that can limit the performance of an 802.11 WLAN [3]. Interfering devices act like DoS attack that prevents an 802.11 radio from transmitting. If there exists another source with strong amplitude greater than 802.11 radios, an 802.11 can sense the energy during the Clear Channel Assessment (CCA) [4] and postpone transmission entirely. The other complicated result of interference is that 802.11 frame become degraded during transmission. These corrupted frames due to interference are retransmitted and will reduce the throughput [5].

Since interference is the core performance-limiting factor in most wireless networks, it is an essential to characterize the interference statistics [6]. Interference has a direct correlation with the Quality of Service (QoS) [7].

## 1.1 Interference and Transmission Range

The zone where a distributing or sending node can be interrupted from a third node during transmission is called interference range. Interference ranges can intensely affect the throughput in wireless sensor networks, due to collisions leading to outage [8]. The interference range finds the area in which other nodes can be interrupted from successful receiving or transmission signals [9].

Xu et al. [10] further indicate that "the interference range can be modeled as a function of distances between source and destination and the physical layer characteristics."

While the transmission range defines the maximum physical range of a radio signals [11]. A transmission range forms the maximum range where a signal can be correctly received. In many wireless sensor network models, transmission range and interference range are supposed to match, meaning nodes cannot interrupt each other when not in transmission range [12]. However, the interference range is bigger than transmission range, depending upon the physical layer modulation scheme. This made interference as interested topic to be investigated further [13].

## 2 CLASSIFICATIONS OF INTERFERENCE

There are several different types of interferences, as described in the following:

### 2.1 Narrowband Interference

A narrowband signal consists of a smaller and limited frequency space and will not cause a Denial of Service (DoS) for an entire band, such as the 2.4 GHz ISM band [14]. A narrowband signal is usually very high amplitude and will absolutely disrupt communications in the frequency space in which it is being transmitted. Narrowband signals can disrupt one or several 802.11 channels. Narrowband interference can also result in corrupted frames and layer 2 retransmissions [15].

### 2.2 Wideband Interference
A source of interference is typically considered wideband if the transmitting signal has the capability to disrupt the communications of an entire frequency band. Wideband jammers exist that can create a complete DoS for the 2.4 GHz ISM band [16].

### 2.3 Multipath
Multipath can cause Inter Symbol Interference (ISI), which causes data corruption. Because of the difference in time between the primary signal and the reflected signals, known as the delay spread, the receiver can have problems demodulating the radio signal's information. The delay spread time differential results in corrupted data. If the data is corrupted because of multipath, layer 2 retransmissions will occur [17].

### 2.4 Adjacent Channel Interference
Most Wi-Fi vendors use the term adjacent channel interference to refer to degradation of performance resulting from overlapping frequency space that occurs due to an improper channel reuse design. In the WLAN industry, an adjacent channel is considered to be the next or previous numbered channel. For example, channel 3 is adjacent to channel 2 [18].

## 2.5 Low SNR

The signal-to-noise ratio (SNR) is an important value because if the background noise is too close to the received signal or the received signal level is too low, data can be corrupted and retransmissions will increase. The SNR is not actually a ratio. It is simply the difference in decibels between the received signal and the background noise (noise floor) [19]. Data transmissions can become corrupted with a very low SNR. If the amplitude of the noise floor is too close to the amplitude of the received signal, data corruption will occur and result in layer 2 retransmissions [20].

## 2.6 Mismatched Power Settings

Another potential cause of layer 2 retransmissions is mismatched transmit power settings between an access point and a client radio. Communications can break down if the client transmit power level is less than the transmit power level of the access point. As a client moves to the outer edges of the coverage cell, the client can "hear" the AP; however, the AP cannot "hear" the client [21].

## 2.7 Near/Far

Disproportionate transmit power settings between multiple clients may also cause communication problems within a Basic Service Set (BSS). A low-powered client station that is at a great distance from the access point could become an unheard client if other high-powered stations are very close to that access point. The transmissions of the high-powered stations could raise the noise floor near the AP to a higher level [22]. The higher noise floor would corrupt the far station's incoming frame transmissions and prevent this lower-powered station from being heard.

## 2.8 Hidden Node

Physical carrier sense and the Clear Channel Assessment (CCA) involves listening for 802.11 RF transmissions at the physical layer; the medium must be clear before a station can transmit. The problem with physical carrier sense is that all stations may not be able to hear each other [23]. Remember that the medium is half-duplex and, at any given time, only one radio can be transmitting. What would happen, however, if one client station that was about to transmit performed a CCA but did not hear another station that was already transmitting? If the station that was about to transmit did not detect any signal during its CCA, it would transmit. The problem is that you then have two stations transmitting at the same time. The end result is a collision, and the frames will become corrupted [24]. The frames will have to be retransmitted. The hidden node problem occurs when one client station's transmissions are heard by the Access Point but are not heard by any or all of the other client stations in the Basic Service Set (BSS).

The clients would not hear each other and therefore could transmit at the same time. Although the access point would hear both transmissions, because two client radios are transmitting at the same time on the same frequency, the incoming client transmissions would be corrupted [25].

Figure 1 shows the coverage area of an access point. Note that a thick block wall resides between one client station and all of the other client stations that are associated to the access point. The transmissions of the alone station on the other side of the wall cannot be heard by all of the other 802.11 client stations, even though all the stations can hear the AP. That unheard station is the hidden node. What keeps occurring is that every time the hidden node transmits, another station is also transmitting and a collision occurs [26]. The hidden node continues to have collisions with the transmissions from all the other stations that cannot hear it during the clear channel assessment. The collisions continue on a regular basis and so do the layer 2 retransmissions, with the final result being a decrease in throughput. A hidden node can drive retransmission rates above 15 to 20 percent or even higher. Retransmissions, of course, will affect throughput, latency, and jitter.
The hidden node problem may exist for several reasons for example, poor WLAN design or obstructions such as a newly constructed wall or a newly installed bookcase. A user moving behind some sort of obstruction can cause a hidden node problem. VoWiFi phones and other mobile Wi-Fi devices often become hidden nodes because users take the mobile device into quiet corners or areas where the signal of the phone cannot be heard by other client stations. Users with wireless desktops often place their device underneath a metal desk and effectively transform the desktop radio into an unheard hidden node [27].

The hidden node problem can also occur when two client stations are at opposite ends of an RF coverage cell and they cannot hear each other, as shown in Figure 2. This often happens when coverage cells are too large as a result of the access point's radio transmitting at too much power. You will learn that an often recommended practice is to disable the data rates of 1 and 2 Mbps on the 2.4 GHz radio of an access point for capacity purposes. Another reason for disabling those data rates is that a 1 and 2 Mbps coverage cell at 2.4 GHz can be quite large and often results in hidden nodes. If hidden node problems occur in a network planned for coverage, then RTS/CTS may be needed.

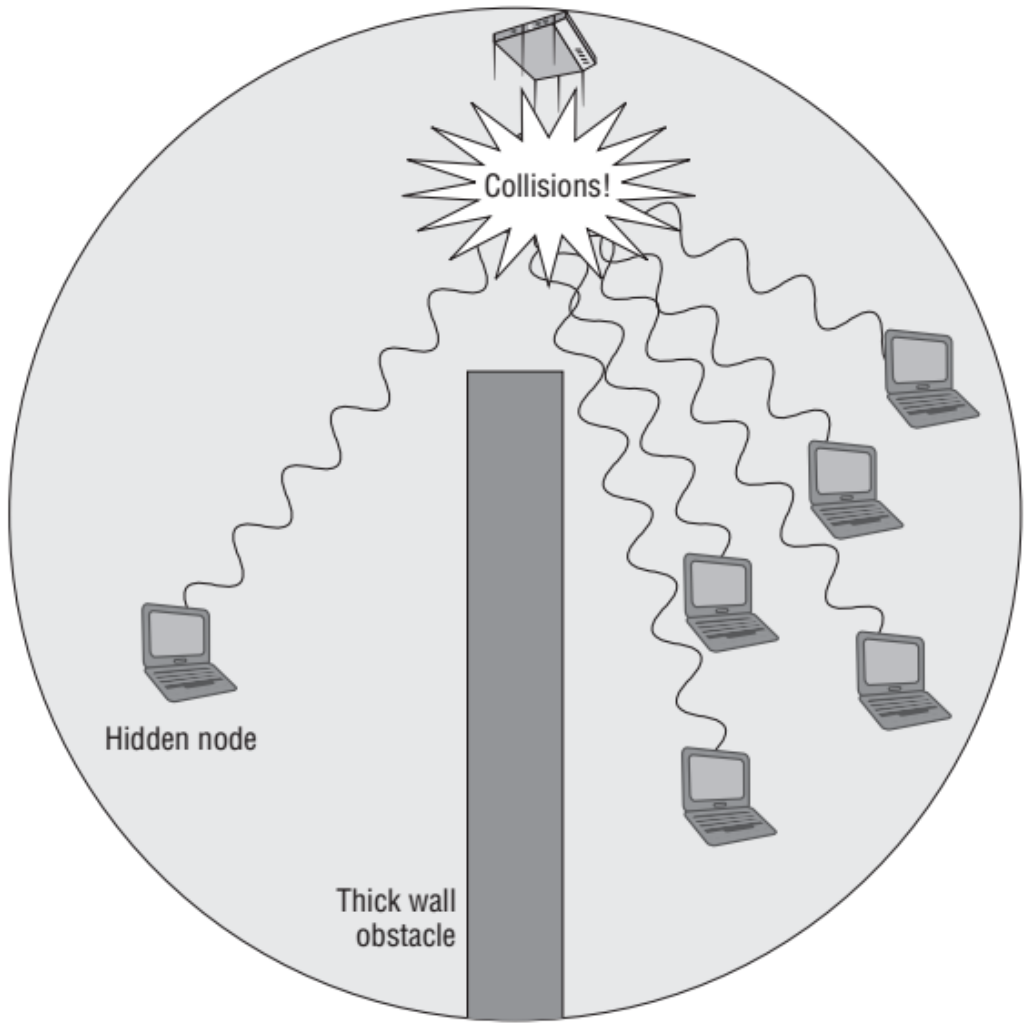

**FIGURE 1** Hidden node—obstruction

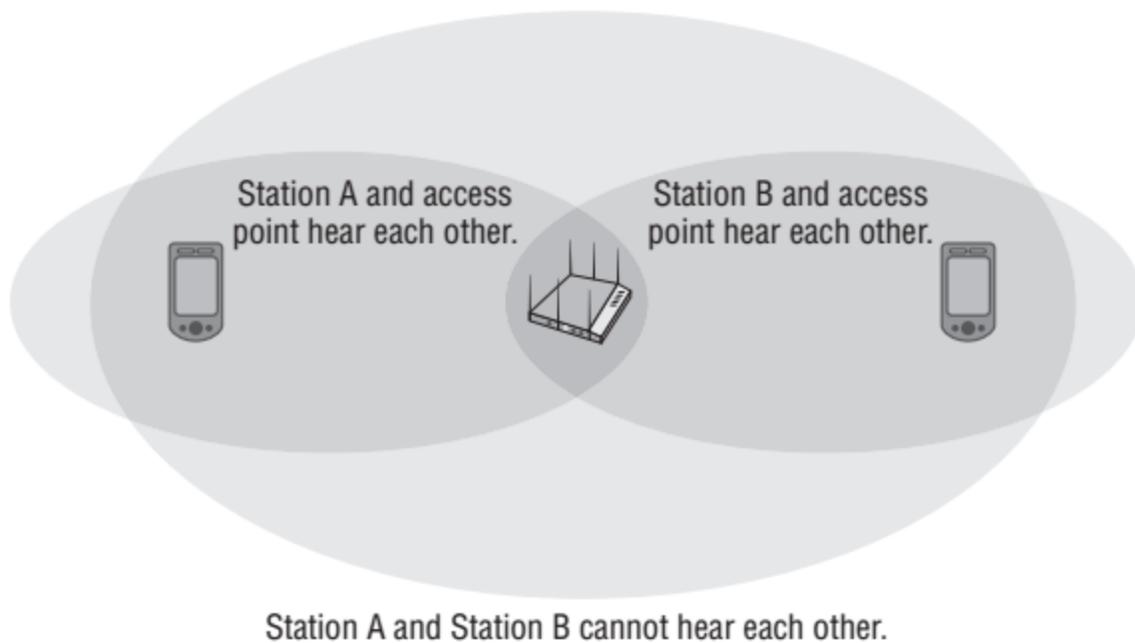

**FIGURE 2** Hidden node—large coverage cell

## 3 RECOGNIZING THE INTERFERENCE

The frequency of an interfering signal is the most common parameter leading to the identification of the interfering source. Thus, an interference problem can often be categorized by its frequency characteristics [28].

### 3.1 Spectrum Analyzer Parameters

What do we need to know about a spectrum analyzer to make sure that we can measure the signal environment adequately? Very basically, we need to know the frequency range, sensitivity, dynamic range, frequency resolution and accuracy.

**3.2 Frequency Range** – Frequency range should be the easiest criteria, since you have a good idea of your system's frequency band and hence the spectrum span you want to observe. Just be sure to give yourself plenty of display width to work with, by setting the frequency span wide enough to include both your affected receiver signals and adjacent interfering signals.

**3.3 Sensitivity** – Sensitivity, although fairly straightforward, can be somewhat confusing. The key is to understand your system specifications and the level of sensitivity required to make your measurements of expected receiver inputs. For example, if your system's receiver signal strength specification is expected to be on the order of –60 dBm, then you will typically only need an additional 20 to 30 dB of measurement range. Thus, a spectrum analyzer that exhibits a sensitivity of –80 to –90 dBm should do the job nicely.

**3.4 Frequency Resolution, Dynamic Range, and Sweep Time** – Frequency resolution, dynamic range, and sweep time are inter-related. Think of resolution as the shape of a scanning "window" which sweeps across an unknown band of signals.

**3.5 Selectivity** – In some interference applications, there will be signals that have amplitudes that are quite unequal. In this case, "selectivity" becomes important criteria. It is very possible for the smaller of the two signals to become buried under the filter skirt of the larger signal [29].

**Shape Factor** – A spectrum analyzer's shape factor defines the ratio of the –60 dB bandwidth to the –3 dB bandwidth of the IF amplifiers.

**Accuracy** – The measurement accuracy of any spectrum analyzer results from the addition of many different accuracy components. Measurement accuracy is important when comparing measured values on unknown signals to published specifications of a system under test. Luckily, when making typical interference measurements, the user is looking for ratios, such as C/I, which determines the operating margin of the desired carrier over the interfering signal in the same operating bandwidth. Thus, absolute accuracy is less critical than relative accuracy.

## 4 CONCLUSIONS AND FUTURE WORK

This article provides the fundamental understanding of interference in Wireless Network, more specifically about WLAN. It elaborates the different used cases of interference and provides a systematic taxonomy about classification and recognitions of interference. As a future work we are devising interference-aware MAC protocol to mitigates the effects of interference.